\documentclass[preprint,showpacs,preprintnumbers,amsmath,amssymb]{revtex4}

\usepackage{graphicx}
\usepackage{textcomp}
\usepackage{mathptmx} 
\usepackage{dcolumn}
\usepackage{bm}

\begin{document}

\preprint{APS/123-QED}

\title{Reducing quantum-regime dielectric loss of silicon nitride for superconducting
quantum circuits}
\author{Hanhee Paik }

\thanks{Present address: Department of Applied Physics, Yale University,
New Haven, CT 06520 }

\email{hanhee.paik@yale.edu}

\author{Kevin D. Osborn}

\affiliation{Laboratory for Physical Sciences \\
 College Park, MD 20740}

\date{\today}

\begin{abstract}
The loss of amorphous hydrogenated silicon nitride (a-SiN$_{x}$:H)
is measured at 30 mK and 5 GHz using a superconducting LC resonator
down to energies where a single-photon is stored, and analyzed with
an independent two-level system (TLS) defect model. Each a-SiN$_{x}$:H
film was deposited with different concentrations
of hydrogen impurities. We find that quantum-regime dielectric loss
tangent $\tan\delta_{0}$ in a-SiN$_{x}$:H is strongly correlated
with N-H impurities, including NH$_{2}$. By slightly reducing $x$ we
are able to reduce $\tan\delta_0$ by approximately a factor of 50,
where the best films show $\tan\delta_0$ $\simeq$ 3 $\times$
10$^{-5}$.
\end{abstract}

\pacs{77.84.Bw, 84.40.Dc, 85.25.-j}

%
\maketitle

Superconducting quantum circuits use amorphous dielectric films for
wiring crossovers and capacitors \cite{Siddiqi05a,Steffen06b,Sillanpaa07a},
but these films often cause loss at low temperatures in the quantum-regime
where the resonator is occupied by a single photon. These defects
can be described by a tunneling two-level system (TLS) model \cite{Schickfus77a}.
There has been recent interest in the low-temperature properties of
dielectrics in superconducting devices \cite{OConnell,Barends,Gao08a}
because they can increase the decoherence rate $1/T_{1}$ in a superconducting
phase qubits \cite{Martinis05a} and the phase noise in microwave
kinetic inductance detectors \cite{Gao07}.

While both films are common in microelectronics, amorphous hydrogenated silicon nitride (a-SiN$_{x}$:H) is found to exhibit less dielectric loss than silicon dioxide (SiO$_{2}$) in the quantum regime \cite{Martinis05a}.  Close to the stoichiometric point $x\simeq4/3$, the electronic and optical properties of this film are determined by the dominant hydrogen impurity which can be changed by deposition conditions \cite{Martinu00a}.

In this paper we first show the composition of five films grown under the condition where there is a strong variation in relative hydrogen impurity type. Then we present low-temperature loss measurements of the a-SiN$_{x}$:H films using superconducting resonators. The loss in the quantum-regime of the films varied dramatically as the relative amount of N-H and Si-H impurities was changed, and in particular we discovered that the NH$_{2}$ concentration is proportional to the TLS density indicating that NH$_{2}$ may be responsible for the microscopic loss. This research demonstrates the low temperature dielectric loss in a-SiN$_{x}$:H can be reduced by reducing nitrogen impurities.

\begin{table*}
\caption{\label{table1} a-SiN$_{x}$:H film and resonator sample properties:
precursor gas ratio $f_{\textrm{N}_{2}/\textrm{SiH}_{4}}$, refractive
index $n$ measured at 633 nm, compressive stress $\sigma_{c}$ (MPa),
ratio of N-H bond concentration to total N-H and Si-H concentration
$c{}_{N-H}/c_{H}$, and five resonator fit parameters.}
\begin{tabular}{ccccccccc}
samples  & $f_{\textrm{N}_{2}/\textrm{SiH}_{4}}$  & $n$ & $\sigma_{c}$ (MPa)  & $c{}_{N-H}/c_{H}$ & tan$\delta_{0} \times 10^6$  & $V{}_{c}$ ($\mu$V)  & $\Delta$  & $f{}_{R}$ (GHz)\tabularnewline
\hline
A (Si rich)  & 1  & 2.159  & 886  & 31 & 25  & 3.8  & 0.35  & 4.970 \tabularnewline
B (Si rich)  & 1.1  & 2.094  & 1178  & 40 & 25  & 2.0  & 0.35  & 4.987 \tabularnewline
C (Si rich)  & 1.12  & 2.065  & 1000  & 38 & 25  & 2.0  & 0.35  & 5.170 \tabularnewline
D (N rich)  & 1.2  & 1.922  & 82  & 72 & 500  & 0.65  & 0.28  & 5.196 \tabularnewline
E (N rich)  & 1.21 & 1.916  & 100  & 83 & 1200  & 0.12  & 0.28  & 5.350 \tabularnewline
\end{tabular}
\end{table*}

The a-SiN$_{x}$:H films were deposited using inductively-coupled
plasma chemical vapor deposition (ICP CVD) in an Oxford Plasmalab
100, using nitrogen (N$_{2}$) and 100\% silane (SiH$_{4}$) precursor
gases. The films were deposited at T = 300 $^{\circ}$C at a pressure
of 5 mTorr, with an ICP power of 500 W, and a nominal rf power of
4 W. Table \ref{table1} summarizes the device parameters and loss
tangent fit results in terms of the precursor gas flow rate ratio of N$_{2}$
to SiH$_{4}$, $f_{\textrm{N}_{2}/\textrm{SiH}_{4}}$. The SiH$_{4}$
flow rate was set to 10 sccm for all films except for E (9 sccm) while
N$_{2}$ flow rate was varied.
\begin{figure}
\includegraphics[width=3.5in]{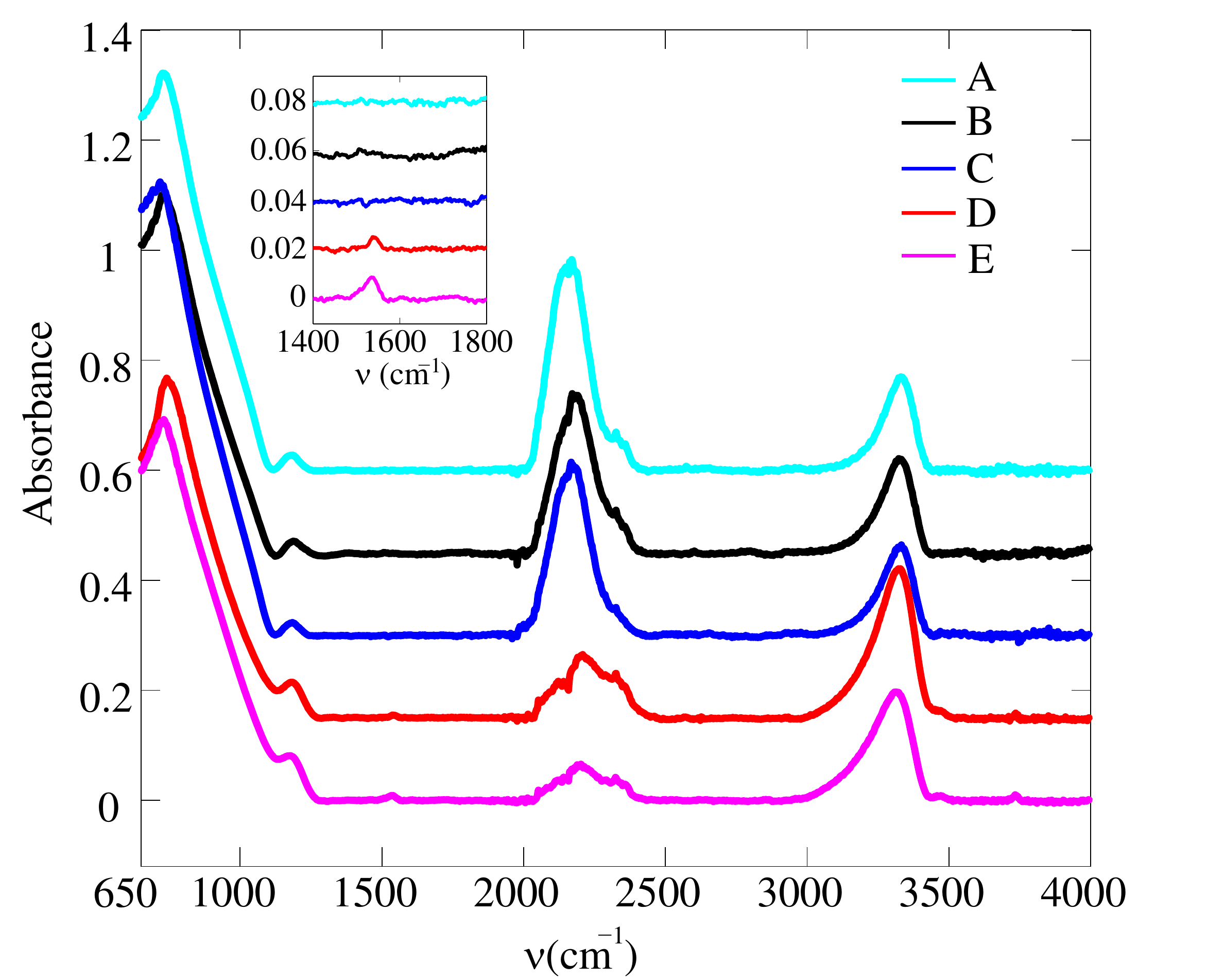}
\caption{\label{FTIR} FT-IR absorbance versus wavenumber $\nu=1/\lambda$,
shown with offsets for better viewing. The nitrogen to silane ratio
increases from film A to E.}
\end{figure}

The refractive index and the compressive stress were used to determine
whether a film is N-rich or Si-rich. \cite{Parsons91a,Martinu00a}.
An N$\&$K Analyzer NKT1500 was used to measure the film refractive
index at $\lambda=$633 nm and the thickness, and a KLA Tencor P-10
profilometer was used to measure film stress. For N-rich films, the
refractive index is below 2 and the compressive stress is significantly
reduced. Only a 10 $\%$ change in $f_{\textrm{N}_{2}/\textrm{SiH}_{4}}$
yielded a 1 GPa change in stress implying a large change in the structure
of the film when the stoichiometry changes between N-rich and Si-rich.
The refractive index and stress were also consistent with a second
set of 5 films that were measured with rf, and the growth rate in the films
also systematically depended on the stoichiometry (not shown). Although
the parameters may depend on the particular machine, a
similar procedure can be performed with another deposition system to find this
crossover stoichiometry.

To evaluate the impurity types of the a-SiN$_{x}$:H films, we measured
FT-IR absorption with a Nicolet 670 attenuated total-reflection FT-IR
spectrometer with a ZnSe prism. The films for FT-IR analysis were deposited to 1 $\mu$m thick to prevent absorption from the underlying substrate and were deposited immediately after those of the resonators.  Figure \ref{FTIR} shows the absorption spectra with baseline correction of five SiN$_{x}$:H films (A to
E). The six absorption bands identified include 1) the Si-N stretching
mode appears at 890 cm$^{-1}$, 2) the N-H bending mode at 1180 cm$^{-1}$,
3) the H-N-H scissoring mode at 1545 cm$^{-1}$, 4) the Si-H stretching
mode at 2180 cm$^{-1}$, 5) the N-H stretching mode at 3340 cm$^{-1}$
and 6) the H-N-H stretching mode at 3460 cm$^{-1}$ \cite{Lucovsky86a,Yin90a,Hanyaloglu98a}.
The relative concentration of N-H bond concentration to the sum of
the N-H and Si-H bond concentrations $c_{N-H}/c_{H}$ is estimated
from the corresponding stretching mode absorbances and the relative
absorptivity of the two modes (shown in Table I). The low value
$(<40\%)$ of $c_{N-H}/c_{H}$ for films A, B, and C and the high
value $(>70\%)$ of D and E indicates that the former films are Si-rich
and the later films are N-rich. It is interesting that NH$_{2}$ absorption modes (both scissoring and stretching) appear strongly in N-rich samples D and E, which shows a N-rich film has a different bonding structure than a Si-rich film as reported previously \cite{Tsu86a,Lucovsky86a}.
\begin{figure}
\includegraphics[width=3.5in]{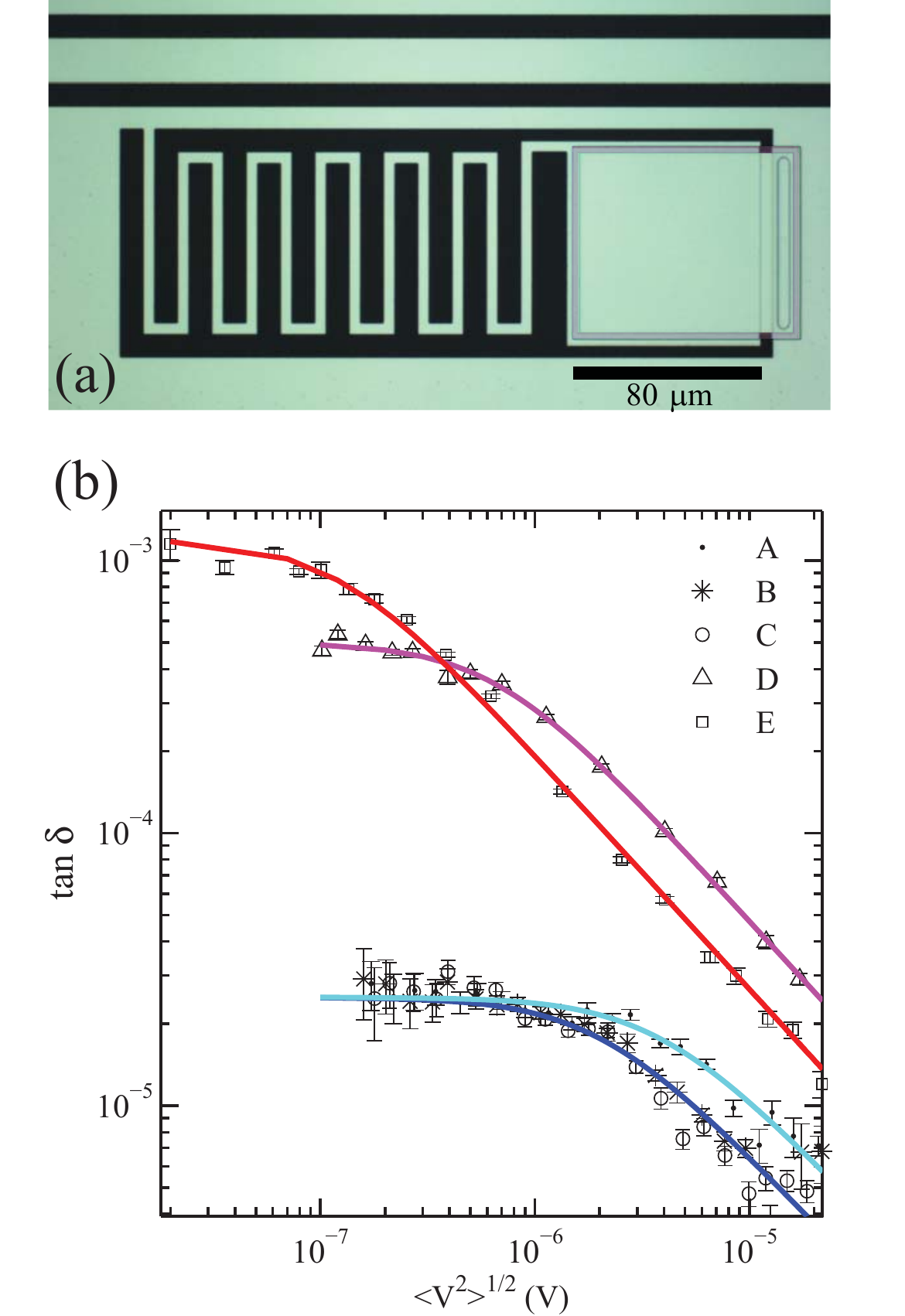}
\caption{\label{circuit} a) Photograph of notch-type aluminum LC resonator
with a SiN$_{x}$ parallel plate capacitor. Size of the capacitor
(shown in right) is 80$\mu$m by 80$\mu$m. The substrate is sapphire
which appears black in the photograph. b) Loss tangent curves of samples
A to E measured at 30 mK (shown with markers) with the two-level system
model fit in Eq. (2) (shown with fit curves).}
\end{figure}

We measured dielectric loss of a-SiN$_{x}$:H films within fabricated Al superconducting LC resonators (see Figure \ref{circuit}). The LC resonator consists
of a meandering inductor $L$ and a parallel-plate capacitor $\hat{C}$
in parallel. Here $\hat{C}=C(1-i\tan\delta)$ where $C$ is the real
part of the capacitance and tan$\delta$ is the loss tangent from
dielectric. The $L$ and $C$ couple to a coplanar waveguide [Figure
2(a)] to form a notch-filter resonator.

For loss tangent measurement, we used an Anritsu 68369A microwave
synthesizer, frequency locked to an Agilent E4440A spectrum analyzer.
The driving line is attenuated with 20 dB attenuators at both 1 K
and the base temperature stages and the input line is calibrated at
room temperature. The return line is isolated by a PAMTech circulator at the 1
K stage and the transmitted signal is amplified with a 4-12 GHz Caltech
HEMT amplifier at 4K. Thermal photons from the circulator at 1 K,
can create on the order of one thermal photon in our resonators, allowing
us to measure coherent response down to the quantum-regime. We obtained
the loaded quality factor $Q$ and the internal quality factor $Q_{i}$
by fitting the measured power transmission to our model function $|t|^2$ given by
\begin{equation}
|t|^2=\left|1-e^{i\phi}\frac{1-Q/Q_{i}}{1+2iQ(f-f_{0})/f_{0}}\right|^{2}.
\end{equation}
A loaded $Q$ and an internal $Q_{i}$ are the fitting parameters
together with a resonance frequency $f_{0}$ and a loss tangent is
given as tan$\delta=1/Q_{i}$. The coupling quality factor $Q_e$ is $\left(Q^{-1}-Q_{i}^{-1}\right)^{-1} \simeq 20000$
for our resonators, and the parameter $\phi$ accounts for small impedance
mismatches for waveguides near our resonator. The capacitor dielectric
has a thickness of 250 nm, and a nominal capacitance of 1.47 pF assuming
a relative permitivity of 6.5, which at $<V^{2}>^{1/2}\approx3\times10^{-7}$
corresponds to an average photon number of $\langle n\rangle\approx$
0.1 (not including thermal photons). However, from the the resonant frequency in Table I, it is apparent that the dielectric constant changes with stoichiometry.

Figure 2(b) shows the loss tangent of a-SiN$_{x}$:H films at 30 mK
as a function of RMS voltage $V$ on the resonator. The error bars are from our $\chi^{2}$ fit of
the resonance peaks to the model function $|t|^2$. The horizontal scale
error is limited by a calibration of our input line at room temperature.
We fit the low temperature loss tangent data to a two-level system
defect model with a parallel-plate geometry which is given as
\begin{equation}
\tan\delta=\frac{\pi\rho(er)^{2}\textrm{tanh}(\hbar\omega/2k_{B}T)}{3\epsilon\sqrt{1+(\Omega_{R})^{2}T_{1}T_{2}}}\simeq\frac{tan\delta_{0}}{\sqrt{1+(V/V_{c})^{2-\Delta}}}
\end{equation}
where $\rho$ is the TLS density of states, $e$ is the electron charge,
$r$ is a distance between two level sites, $\varepsilon$ is a permittivity
of the film, $\Omega_{R}=eVr/\hbar d$ is a TLS Rabi frequency, $V$
is an RMS voltage, d is the thickness of a-SiN$_{x}$:H, $\omega$ is the angular resonance frequency and $T$ is the temperature of the resonator \cite{Martinis05a}.
Here the intrinsic loss tangent tan$\delta_{0}$ which is defined as the dielectric loss in a single photon regime is a function of TLS density of states $\rho$ and $V_c$ is a threshold voltage where saturation starts to occur in TLS \cite{Phillips87a}.  The fitting parameters are summarized in Table I.

Surprisingly, Si-rich SiN$_{x}$:H films (A, B and C) showed a 20 and 50 lower intrinsic loss than the N-rich films (D and E). This also implies that the TLS density of states $\rho$ is correlated
with the N-H bond concentration. We believe the results are reproducible
from the fact that the samples A, B (low loss) and D (high loss) were
fabricated 8 months after the samples C (low loss) and E (high loss)
were made. We confirmed our film's reproducibility by making
another low-loss a-SiN$_{x}$:H film 4 months after these films A,
B, and D, using the same recipe as the sample A. With that new film
we obtained tan$\delta_{0}\simeq$ 3$\times$ 10$^{-5}$, which
agrees with the original measurement of film A.

\begin{figure}
\includegraphics[width=3in]{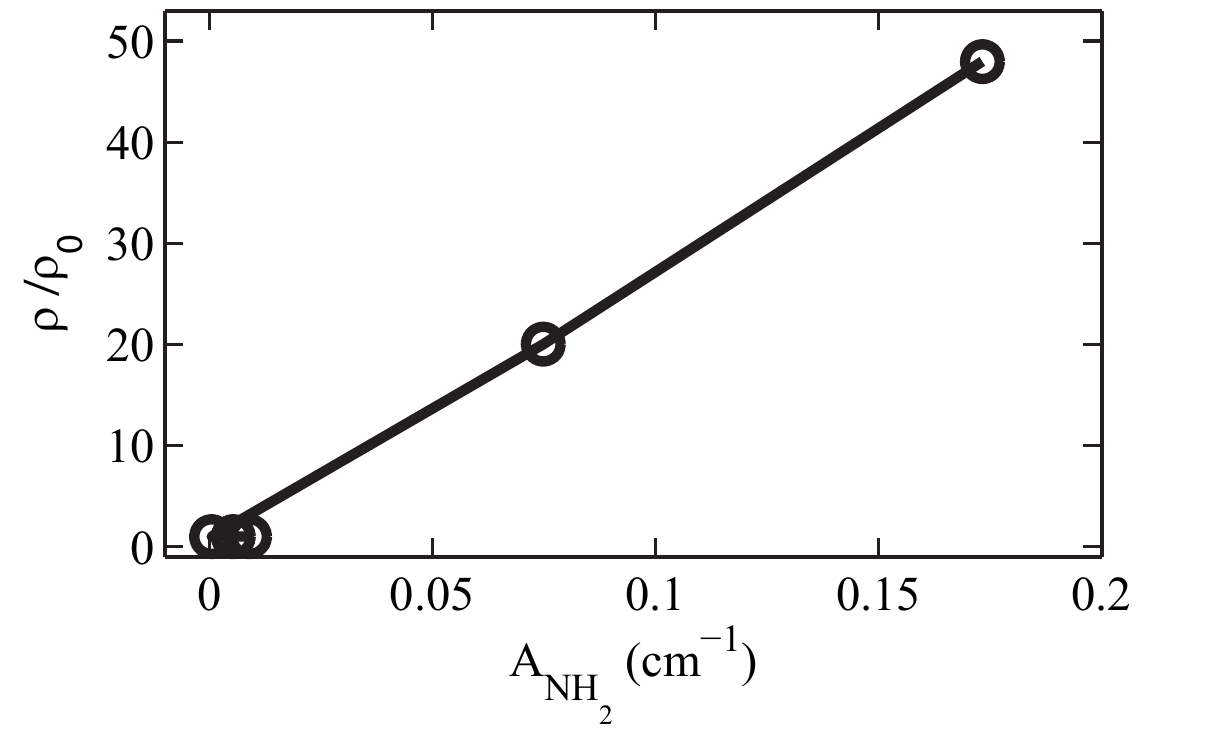}
\caption{\label{dos} Density of states as a function of NH$_{2}$ bonding
peak area in FT-IR measurement.}
\end{figure}

In Figure \ref{dos}, we plot $\rho$ normalized by the TLS density of
states $\rho_{0}$ of the lowest loss film (from samples A, B and
C) as a function of NH$_{2}$ scissoring mode absorption band area
A$_{NH_{2}}$, which is proportional to the concentration of the NH$_{2}$
bond \cite{Lanford78a,Yin90a,Parsons91a}.  We find that $\rho$ is proportional
to A$_{NH_{2}}$ and the amount of NH$_{2}$ in the three Si-rich films (samples A, B and C) is much
smaller than the N-rich samples. (Note that the NH$_2$ stretching mode almost disappears for the Si-rich
films A, B and C.) Therefore, for the two N-rich a-SiN$_{x}$:H films, NH$_{2}$
is likely responsible for the TLS dielectric loss.

The mechanism by which NH$_{2}$ bonding causes low temperature loss can be understood by comparing the film with one composed of silicon dioxide with hydrogen impurities (SiO$_{x}$:H). The N-rich a-SiN$_{x}$:H is known to have a local bonding arrangement that is usually seen in Si(NH)$_{2}$ \cite{Tsu86a,Lucovsky86a} which is isoelectric and isostructural with SiO$_{x}$:H.  In this case, $O$ is substituted with $NH$ \cite{Tsu86a,Lucovsky86a,Martinu00a}, thus the NH$_{2}$ impurity in a-SiN$_{x}$:H is analogous to the OH impurity in SiO$_{x}$:H.  The molecular motion of the OH bond is believed to cause the low-temperature loss in a-SiO$_{2}$ \cite{Schickfus77a,Phillips87a}; and similarly, the molecular motion of NH$_{2}$ may cause the loss in a-SiN$_{x}$:H.

The measured loss tangent includes the loss from the few nanometer
thick native aluminum oxide on top of the bottom Al electrode, which
was not removed. In our case with a parallel plate capacitor, the
field energy that resides in the native oxide is on the order of $10^{-2}$
of the total electric field energy. If we assume that the observed
loss is entirely from the native oxide, then the loss tangent of the
aluminum oxide is $\tan\delta_{0,AlOx}\simeq$ 3$\times$10$^{-3}$,
which is similar to the previously reported estimates \cite{Martinis05a}.

In conclusion, we have measured the quantum-regime loss tangent of
five a-SiN$_{x}$:H films at 30 mK. Each film was grown with a different
precursor gas flow rate ratio $f_{\textrm{N}_{2}/\textrm{SiH}_{4}}$, such that films were rich in either Si-H or N-H impurities. We found that the NH$_{2}$ bond in
N-rich a-SiN$_{x}$:H is correlated with TLS-induced dielectric loss
and were able to reduce the quantum-regime loss tangent to 3 $\times$
10$^{-5}$ by making a-SiN$_{x}$:H films Si-rich.

The authors thank John Martinis, Ben Palmer, Dave Schuster, and Fred
Wellstood for useful discussions, and Dan Hinkle for his advice on
amorphous silicon nitride deposition. This work was funded by the
National Security Agency.


\begin{thebibliography}{10}


\bibitem{Siddiqi05a} I. Siddiqi \textit{et al.}, Phys.\ Rev.\ Lett.
\textbf{94}, 027005 (2005).

\bibitem{Steffen06b} M. Steffen \textit{et al.}, Phys.\ Rev.\ Lett.
\textbf{97}, 050502 (2006).

\bibitem{Sillanpaa07a} M. A. Sillanpaa \textit{et al.}, Nature \textbf{449},
438-442 (2007).

\bibitem{Schickfus77a} M. von Schickfus and S. Hucklinger, Phys.\ Lett.
\textbf{64A}, 14 (1977).

\bibitem{Phillips87a} W. A. Phillips, Rep.\ Prog.\ Phys. \textbf{50},
1657 (1987).

\bibitem{OConnell} Aaron D. O'Connell\textit{ et al.}, APL \textbf{92},
112903 (2008).

\bibitem{Gao08a} J. Gao \textit{et al.}, Appl.\ Phys.\ Lett. \textbf{92},
152505 (2008).

\bibitem{Gao07} J. Gao \textit{et al.}, Appl.\ Phys.\ Lett. \textbf{90},
102507 (2007).

\bibitem{Barends} R. Barends\textit{ et al.}, Appl.\ Phys.\ Lett.
\textbf{92}, 223502 (2008).

\bibitem{Martinis05a} J. M. Martinis \textit{et al.}, Phys.\ Rev.\ Lett.
\textbf{95}, 210503 (2005).

\bibitem{Martinu00a} L. Martinu and D. Poitras, J.\ Vac.\ Sci.\ Technol.\ A
\textbf{18}, 2619 (2000).

\bibitem{Lucovsky86a} G. Lucovsky and D. V. Tsu, J.\ Vac.\ Sci.\ Technol.\ A
\textbf{4}, 681 (1986).

\bibitem{Yin90a} Z. Yin and F. W. Smith, Phys.\ Rev.\ B \textbf{42},
3666 (1990).

\bibitem{Hanyaloglu98a} B. F. Hanyaloglu and E. S. Aydil, J.\ Vac.\ Technol.\ A
\textbf{16}, 2794 (1998).

\bibitem{Tsu86a} D. V. Tsu \textit{et al.}, Phys.\ Rev.\ B \textbf{33},
7069 (1986).

\bibitem{Parsons91a} G. N. Parsons \textit{et al.}, J.\ Appl.\ Phys.
\textbf{70}, 1553 (1991).

\bibitem{Lanford78a} W. A. Lanford and M. J. Rand, J.\ Appl.\ Phys.
\textbf{49}, 2473 (1978).


\end{thebibliography}
\end{document}